# Transaction Fraud Detection Using GRU-centered Sandwich-structured Model

Xurui Li[1, 2], Wei Yu[3], Tianyu Luwang[4], Jianbin Zheng[1], Xuetao Qiu[1], Jintao Zhao[1], Lei Xia[3], Yujiao Li[4]

*Research Institute of Electronic Payment, China UnionPay[1], Shanghai, China*
*School of Computer Science, Fudan University[2], Shanghai, China*
*Artificial Intelligence TSS Team, Intel Corporation[3], Shanghai, China*
*ZhongAn Information Technology Services Co.,Ltd.[4], Shanghai, China*
*\*Corresponding author: xurui.lee@msn.com;*

*Abstract*—Rapid growth of modern technologies is bringing dramatically increased e-commerce payments, as well as the explosion in transaction fraud. Many data mining methods have been proposed for fraud detection. Nevertheless, there is always a contradiction that most methods are irrelevant to transaction sequence, yet sequence-related methods usually cannot learn information at single-transaction level well. In this paper, a new "within→between→within" sandwich-structured sequence learning architecture has been proposed by stacking an ensemble model, a deep sequential learning model and another top-layer ensemble classifier in proper order. Moreover, attention mechanism has also been introduced in to further improve performance. Models in this structure have been manifested to be very efficient in scenarios like fraud detection, where the information sequence is made up of vectors with complex interconnected features.

## I. INTRODUCTION

Occurrence of fraudulent transactions has been growing rapidly with the booming development of e-commerce. It costs consumers and financial institutions billions of dollars annually. According to the Nilson Report, global card fraud cost in 2016 has reached $22.80 billion. The fraud rate in 2016 has reached 7.15 BP (1 BP = 0.01%), which increased by 60% compare to that of 2010 [1]. Therefore, fraud detection has become the vital activity to reduce fraud impact on service quality, costs and reputation of a company or institute. Traditional anti-fraud method relying on manual audit is unable to deal with explosively growing information data. Rule engines have already been widely involved in many transaction systems. Meanwhile, criminals are keeping on finding new tricks by avoiding known rules to commit fraud actions. It makes rule-based fraud detection method hard to handle ever-changing fraud patterns. Consequently, financial institutions are struggling to find more intelligent methods for detecting fraud events, with the goal of reducing fraud losses as much as possible.

## II. RELATED WORK

Machine learning methods have already been introduced into fraud detection area. Supervised models like logistic regression (LR), support vector machine (SVM) and random forest (RF) are estimated by labeled historical transaction data [2, 3]. They use the trained model to predict whether a new transaction is fraudulent or legitimate. Unsupervised methods like isolation forest (IF) usually identify outliers as potential fraudulent cases [4]. It can help detect some new fraud patterns which have not been found previously. Nevertheless, most of these methods treat each transaction as an independent individual and ignore the associations between them. However, these sequence-related factors may have significant influences on the outcome of fraud detection model. For instance, criminals may try some small amount tentative deals before carrying out large amount transactions. Some of these patterns can be artificially calculated as candidate features, but it depends too much on expert experiences and lacks duly comprehensive consideration.

Some behavior-based algorithms such as hidden markov model (HMM) and peer group analysis (PGA) have been proposed for fraud detection by discovering anomalies comparing to regular transaction patterns of an account [5, 6]. However, most behavior-based models need to be constructed separately for each account. It relies on account's exact historical regular patterns, which are difficult to obtain. Recently, deep learning methods based on recurrent neural networks (RNN) have been proved to be with good performance in sequence analysis work [7]. Dynamic temporal behaviors for various accounts can be analyzed with help of sequence labeling skills by RNN [8]. Nevertheless, just as most sequence analysis methods, although more sequential information between transactions can be extracted, the feature learning ability within a single transaction is insufficient for RNN models. These relationships within a single transaction can be well learned by some classification models like RF, but at the expense of attenuating the sequential learning ability.

In this paper, a comprehensive building process for transaction fraud detection model with the collaboration of various platforms and algorithms has been presented. A new "within→between→within" (WBW) sandwich-structured sequence learning architecture has been proposed by combining ensemble and deep learning methods. Models in similar structures will show exciting performances particularly in scenarios like transaction fraud detection, where the sequence is made up of vectors with complex interconnected features.

Our paper is organized as follows. The overall idea for building model is elaborated in Section III. Firstly, feature engineering work with rich expert experiences is done on distributed computing platforms for massive transactions. Then detailed construction of WBW architecture is described. Attention mechanism has also been involved for enhancing model performance. The whole model has been validated on actual transaction data of UnionPay at counterfeit credit card fraud detection scene, with the effect manifested in section IV. At last, we conclude the paper in Section V.

## III. DESIGN AND IMPLEMENTATION

### A. Artificial Feature Engineering

Each transaction should first be mapped into a row vector based on original transaction fields. Additional feature engineering is necessary for these vectors. According to our previous experiences, more derivative variables can be calculated using statistical methods combing some skills like rolling-window and recency-frequency-monetary (RFM) framework [9]. For example, the amount for current deal, last deal and the variance between them should both be treated as features for current transaction. Total amount over different time periods are computed as different features. Some trusted characteristics can also be used as features according to our analysis or blacklists. For instance, if many fraud cases happen in a specific location according to historical transactions, then the customized feature "*is_high_risk_loc*" for this location is 1, otherwise 0.

Some continuous numerical variables like money amount and transaction moment can be directly accepted by the following model, while further artificial analysis on them can enhance the efficiency. We can use weight of evidence (WOE) method to discretize them into more distinguishable features. By the way, with the help of Spark libraries like ML or MLib, the quality of features can be further improved. For instance, the "*VectorIndexer*" API can automatically identify categorical features like location and merchant type and index them into discrete numeric variables. Moreover, the "*StringIndexer*" API not only discretize categorical features, but also order them by frequencies, which help improve the model performance commendably.

### B. Feature Optimization Based on GBDT

Now each transaction is mapped into a vector at $n_A$ dimension. The above experience-based feature engineering work is indispensable, because it can help the subsequent model to learn the inherent characteristics more quickly and accurately. However, artificial experience still encounters omissions inevitably. For example, an off-site transaction with large amount happened at midnight may be very suspicious, while the separate "off-site", "large-amount" and "midnight" are all common features. There are many similar effective combinations of features, and some of them are hard to be found artificially. To make up for the lack of manual experiences, researchers from FaceBook have tried to use gradient boost decision tree (GBDT) model to help discover latent combinations between features automatically before using LR classifier [10]. GBDT preferred generating features with more overall discriminabiltity before generating distinguishing features for few samples. It is why they choose GBDT feature learning rather than RF. In addition, the trained GBDT model needs to be saved for feature conversion during subsequent fraud detection phases.

After the transaction been transformed by GBDT into a vector at $n_G$ dimension, we concentrate the original vector of $n_A$ dimension with the GBDT vector into a new vector $V_{sg}$, with the dimension at $n = n_A + n_G$. The reason why we use the concentrated vector of n dimension instead using GBDT vector directly for the following transformations is because more deep information could be obtained if original features are also involved in sequential learning using the gated recurrent unit (GRU) model.

### C. Sequential Features Learning Based on RNN

a) Introduction of GRU model

Traditional machine learning methods cannot handle relationships between transactions well. RNN is a kind of neural networks who maintain a hidden state which can remember certain aspects of the sequence it has seen. However, the magnitude of weights in transition matrix can have strong impacts on the learning process during gradient back-propagation phase of basic RNN, which may lead to situations called vanishing or exploding gradients. Long short term memory (LSTM) model introduces a new structure called memory cell to enable the capability of learning long-term dependencies [11]. On top of this, the GRU model modifies 3 gates (input/forget/output) of LSTM into 2 gates (update/reset), and merges the unit\output into one state [12]. It reduces the matrix multiplication to some extent. Thus the GRU model is adopted here considering the relative large training data.

Fig. 1 shows the diagrammatic sketch for GRU. Supposing there are j hidden node units. The formulas for each unit are as follows:

$$r_t^j = \sigma(W_r \mathbf{x_t} + U_r \mathbf{h_{t-1}})^j$$
$$z_t^j = \sigma(W_z \mathbf{x_t} + U_z \mathbf{h_{t-1}})^j$$
$$\hat{n}_t^j = \tanh(W \mathbf{x_t} + U(\mathbf{r_t} \odot \mathbf{h_{t-1}}))^j$$
$$h_t^j = z_t^j h_{t-1}^j + (1 - z_t^j)\hat{n}_t^j$$

Here, $r_t^j$ is reset gate, which allows model to drop information that is irrelevant in the future. $z_t^j$ is update gate, which controls how much of past state should matter now. $\hat{n}_t^j$ is new memory and $h_t^j$ is current hidden state. $\odot$ means element-wise multiplication, $\sigma$ represents sigmoid function. $[.]_j$ means the No. j element for a vector. $W_r$, $U_r$, $W_z$, $U_z$, W and U are matrix weights to be learned. It can be found that the new memory is generated by previous hidden states and current input. Previous hidden state will be ignored when $r_t^j$ is close to 0. $z_j^{(t)}$ determines the influences of $h_{t-1}^j$ and $\hat{n}_t^j$ to current hidden state $h_t^j$.

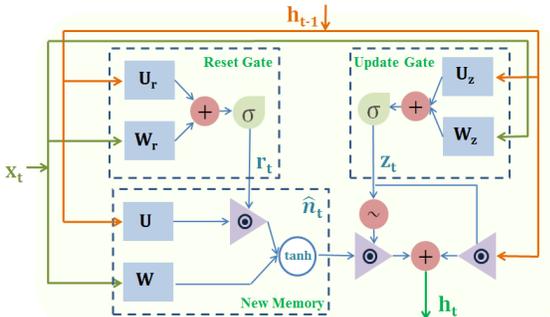

*Fig. 1. Diagram for a single hidden node unit in GRU*

b) Generate sequential samples

After observations been sampled into balanced ratio, samples need to be transformed into sequential ones whose format the GRU model can handle. We extract and group the samples as follows:

1. Group the transactions by account and count the number of transactions for each account.

2. Separate the accounts into different sets according to their transaction counts.

3. For each set i, the transaction times for a single account is in range of $[S_i, E_i]$.

After transaction been grouped into corresponding set, sort the transactions by time for each account belong to set i. Current vector $V_{sg}$ for the No. r transaction now is $X_r = \{x_{r1}, x_{r2}, \dots, x_{rn}\}$. Each transaction can be then extended into a sequential vector sample with fixed dimension at $n * E_i + 1$. It means the parameter "timesteps" (TS) for this set is $E_i$. For the earliest transaction of the account, the front elements are all filled with 0 because there is no previous transaction recorded. By appending the "fraud or normal" label $Y_1$ of current transaction, the first sequential sample can be described as $\{0 \dots 0, X_1, Y_1\}$. For the No. r transaction ($r < E_i$), the previous $r-1$ transactions are arranged before current one, and the sequential sample for this transaction can be described as $\{0 \dots 0, X_{r-1}, X_r, Y_r\}$ by appending the current label $Y_r$. The dimension of elements filled with 0 is $n*(E_i - 2)$. And if r is the last transaction for this account, stop generating samples for it. If the last transaction happens to be $E_i$, the last sequential sample for this account is $\{X_1 \dots, X_{E_i-1}, X_{E_i}, Y_{E_i}\}$, with no elements filled with 0. Typical sample extension process is shown in Fig. 2.

Detail division of TS should also be modified according to actual situations. For instance, a threshold of maximum transaction counts $E_M$ can be defined. Accounts with transaction count exceed $E_M$ can be categorized into the last set $[S_k, E_M]$. And for the No. t transaction ($t > E_M$) of an account, only the previous $E_M - 1$ transactions will be involved to generate the sequential sample using moving window with size of $E_M$, namely $\{X_{t-(E_M-1)} \dots, X_{t-1}, X_t, Y_t\}$.

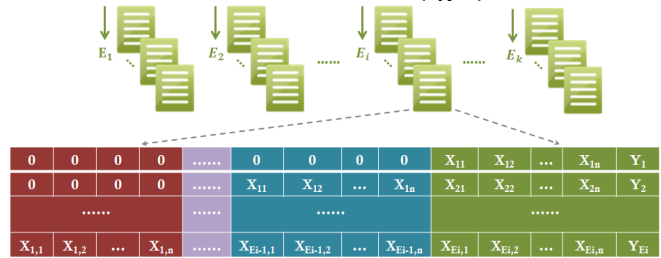

*Fig. 2. Typical sequential sample generation framework*

c) Sequential Features Learning using "WB" Structure

After generating sequential samples, we construct sequential feature optimization model for each of the k sets. GRU model could be trained to detect fraud directly by appending a 2-dimensional output of softmax layer, yet we do not intend to do so. Although RNN model can learn relevance between sequential items well, the learning ability for features inside a single transaction is only equivalent to a basic shallow network. Engineers from Google have raised a "wide & deep learning" framework by combing LR model with basic deep neural networks (DNN) to improve feature learning ability [13]. Analogously, we first proposed a "within & between" (WB) structure by combining GBDT and GRU model to intermingle their advantages. GRU model here is applied for extracting more potential sequential optimized features based on previous GBDT learning process. In fact, a recent literature has already shown a relatively good structure by using the output of LSTM as the input for another classifier like LR. This can be taken as a forerunner for "between & within" (WB) structure [14].

Detail for sequential features learning model in WB structure is shown in Fig. 3. As can be seen, the blue dots represent the $n_A$ dimensional vector after artificial feature engineering, purple dots represent vectors optimized by GBDT. These two vectors are merged into $V_{sg}$ for a single transaction. Sequential samples are reshaped into dimension of $(E_i, n)$, and this is the input of GRU model. Supposing the number of layers is $N_{layer}$ and the output dimension is $n_D$ for GRU. These parameters can be adjusted according to actual demands and effects. Here we take $N_{layer} = 2$ as an example. After GRU model has been trained, the output of the last node could be the sequential feature vector $V_{sq}$, as shown in the inset of Fig. 3. Preferably, a mean pooling layer can be applied on top of the last GRU layer to learn more sequential information among transactions. It means the average result of all $E_i$ nodes is taken as $V_{sq}$. Note that although there are k different sequential optimization models for k sets, the output dimension for each model only depends on the parameter $n_D$ for GRU model. It means the format of sequential features is consistent for all transactions if parameter $n_D$ stays the same. But one fixed parameter $n_D$ may not have good performance for all k models. So it is suggested to append an additional multi-layer perceptron (MLP) on top of current GRU output or the mean pooling layer, with the dimension for the last layer of MLP keep the same value of $n_M$ for all k models. Here, a single-layered dense is also acceptable instead of MLP. Then we can concentrate the $V_{sg}$ of current transaction and sequential feature vector $V_{sq}$ into a new vector with dimension at $n_{op} = n + n_M$, which is the final optimized eigenvectors $V_{op}$ for current transaction.

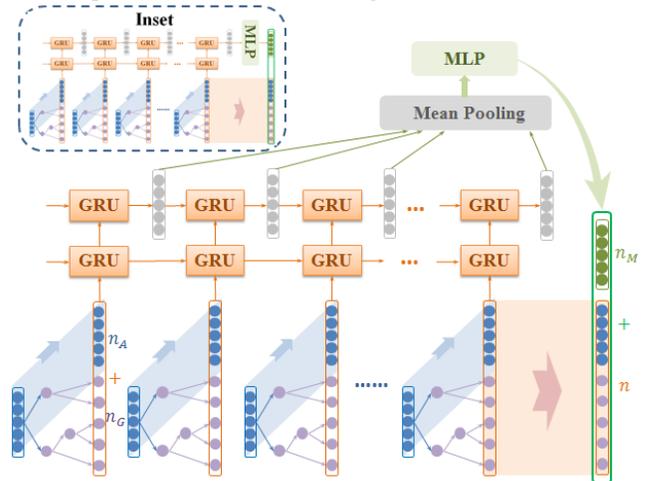

*Fig. 3. Typical WB sequential features learning structure. The insert is a simpler structure without mean pooling layer.*

As a special case, no division work will be done. All transactions will be processed by one sequential model with a same TS parameter at $ts$. It means only previous $ts$ transactions would be taken into consideration when building sequential model for one account even if there are more transactions before. Sequential information can then be learned in a more simple way at the expense of a little approximation loss. Better solution is to introduce the attention mechanism into our GRU model, whose output is weighted average of sequential inputs [15]. A sketch-map for model combined with attention mechanism is shown in Fig. 4. Each GRU cell is paired with one attention model (AM). As can be seen from the insert of Fig. 4 that the AM unit first compute each $m_i$ with a tanh layer as: $m_{ij} = v^T \tanh(W_m h_{i-1} + U_m V_{sgj})$. Here $v^T$, $W_m$ and $U_m$ are learnable parameters. Each weight $s_{ij}$ is computed by softmax function as: $s_{ij} = \exp(m_{ij}) / \sum_{k=1}^{TS} \exp(m_{ik})$. At last, the output $z_i$ can be computed as the weighted average of all $V_{sgi}$: $z_i = \sum_{j=1}^{TS} s_{ij} V_{sgj}$. Note that the architecture illustrated in Fig. 4 is about the feature-level attention [16]. The mean polling layer can also be replaced by another AM, which could be the component-level attention. This work will be tried in future work. Anyway, by using attention mechanism, it is possible to focus on the interesting part of sequences regardless of the size of input sequence. Models with different TS can also include attention mechanism respectively in the same way.

### D. Entire Fraud Detection Model Based on "WBW" Sequence Learning Architecture

Final fraud detection model can then be trained using a top-layer classifier based on the optimized eigenvectors $V_{op}$. The classifier can be chosen from some common algorithms, but suggested ones are ensemble models such as RF, GBDT and extreme gradient boosting (XGBoost), which have been proved to be very effective in fraud detection area. To synthesize the advantages of multi methods, boosting ensemble methods will no longer be selected because GBDT has already been involved in previous feature learning steps. So RF model is selected here as the top-layer classifier, which is one of the best bagging ensemble methods and can be implemented in parallel well.

Let's review the whole model training process. First, the GBDT model is used to expand effective features within single transaction. Then the GRU model is implemented to learn sequential features between transactions. At last, the RF model is applied to relearn more potential features within the optimized eigenvectors $V_{op}$ for each transaction. Similar structures can be unified into a "WBW" sandwich-structured sequence learning architecture. The advantage of the WBW architecture is intuitively interpretable. The features obtained from ensemble models like GBDT could be sequentially dependent between transactions besides artificial calculated features. For example, "a large amount off-site transaction at midnight" would happen after some tentative "small amount off-site transactions at midnight". Similar suspicious fraud patterns in deeper levels with sequential dependencies can be well learned automatically by the first WB structure. Meanwhile, newly learned sequential features again may have potential associations with others within a transaction. For instance, a "current large amount transaction happens after some small tentative ones" pattern would be more suspicious if combined with other patterns like "current transaction location is different from the previous tentative ones". It means the sequential features may be recombined into new features within a single transaction for exposing deeper information. Similar information can be learned intelligently by the second "between→within" (BW) structure. The typical processing flow for entire WBW framework is shown in Fig. 5.

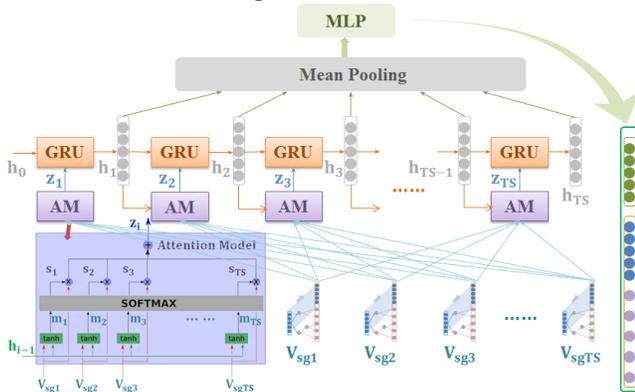

*Fig. 4. Sketch-map of WB structure with attention model. Bottom left inset is a schematic diagram of AM*

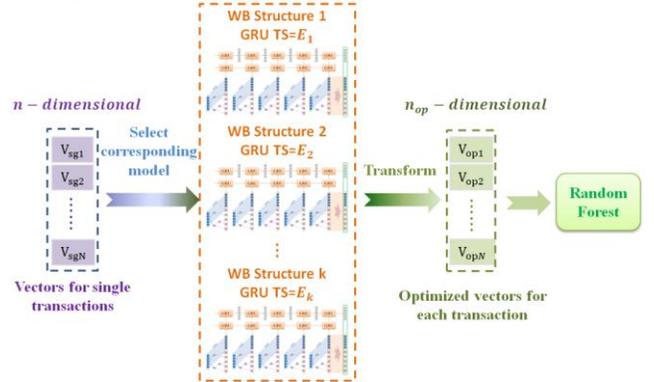

*Fig. 5. Typical processing flow for entire WBW framework*

## IV. EXPERIMENTS AND RESULTS

Our original data are stored in Hive tables among clusters based on Cloudera CDH-5.9.0. Feature engineering is implemented on Spark-2.1.0. The experimental cluster consists of 100 nodes, where each node contains an Intel Xeon CPU E5-2620 at 2.00GHz CPU and 8 GB RAM. Data in Hive tables can be directly read by Spark SQL into Spark *DataSet* for further processing. The GRU related models are carried out on Tensorflow-1.2.0. Spark ML library was tried for the ensemble models like RF and GBDT at first, but the performance is very poor in seriously imbalanced situations. This may be caused by some approximation in the process of algorithm parallelization. So the Scikit-learn library is selected for these ensemble models.

There are also various types of fraud. Taking the statistical result of China in 2016 as an example, the most fraud type for debit cards is telecommunication fraud, while the fraud losses for credit cards are dominated by counterfeit cards. Special models should be built for different fraud types separately. Here we take the counterfeit credit card fraud detection model as an example to show the typical building process and performance. The WBW sequence learning process was trained on a real transaction collection of UnionPay within a three-month period from 2016.06 to 2016.08. The division of parameter TS was implemented according to the distribution of transaction count for single accounts. As shown in Fig. 6, the histogram represents the statics of the accounts number within special transaction count range for a single account during experimental period. The first red column represents that there are about $2 \times 10^5$ accounts whose transaction count is in range of 0~5, while the second green column represents that in range of 5~10. Special sequential model with corresponding TS was built for accounts within these ranges with large accounts number. Whereas accounts in the range of 40~100 was arranged into one sequential model with a uniform TS (as shown in orange color) at 40 due to the relative less accounts number. A special sequential model with TS at 100 was also built for accounts in the range above 100 considering the larger total number of transactions.

Performances of multiple algorithms have been compared. The precision and recall of fraudulent samples are good choices for performance evaluation in view of the highly skewed data. Fig. 7 (a) shows the precision-recall (PR) curve of test data in the following month of 2016.09 for each algorithm with the imbalance ratio between regular and fraudulent samples at 10000:1. As can be seen, the performances for ensemble models like RF and GBDT are better than other common classification models like SVM or LR under the experimental scenario. RF combined with GBDT optimization has only a little promotion compared to separate ones. Single GRU sequential model prevails a little over ensemble models. Some small improvement can be obtained by placing ensemble models before or after GRU process. By contrast, more distinct promotion emerges when GBDT, GRU and RF models are stacked in order. In another word, GBDT→GRU→RF (WBW) model has better performance than that of GBDT+RF→GRU (WWB) or GRU→GBDT+RF (BWW) model. Besides, from Fig. 7 (b) we can see that the predicting ability for RF model attenuates gradually as time elapses, while the effect of GRU model is declining with some irregular beatings. It means that current fraud detection patterns within a single transaction are ever-changing while the sequential patterns could be effective periodically. Nevertheless, it is suggested that all models should be trained termly in case of losing effectiveness.

In fact, RF model could be better than single GRU model when data is balanced. As shown in Fig. 8 (a) the best F1 score is higher than that of GRU model at first, while it drops more sharply with increasing imbalance ratio. It means GRU model can alleviate imbalance to some extent. The WBW approach inherits this advantage of GRU, and can give a relatively good performance in seriously imbalanced situations. The performances of WBW models with different GRU structures have also been compared, as shown in Fig. 8 (b). It can be seen that model with GRU consists of various artificially divided TS ranges has better performance than that with fixed TS. Further improvement can be accessed by combining attention mechanism. It can also be found that models with GRU at larger TS benefit more from attention mechanism, and the performance of model with AM at fixed TS=20 is very close to that of model using various GRU models in different time steps. It indicates that attention-based single GRU model at a relatively large TS is also an acceptable choice for the "B" in WBW structure, considering the simplicity of artificial operation.

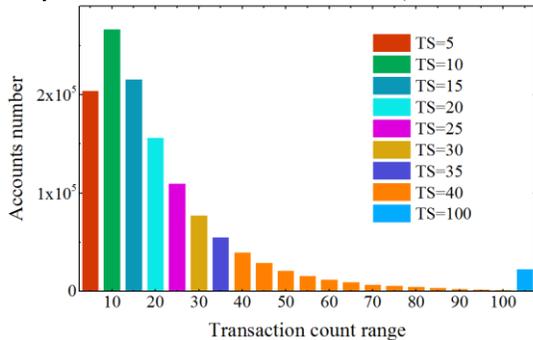

*Fig. 6. Distribution of transaction count for single accounts*

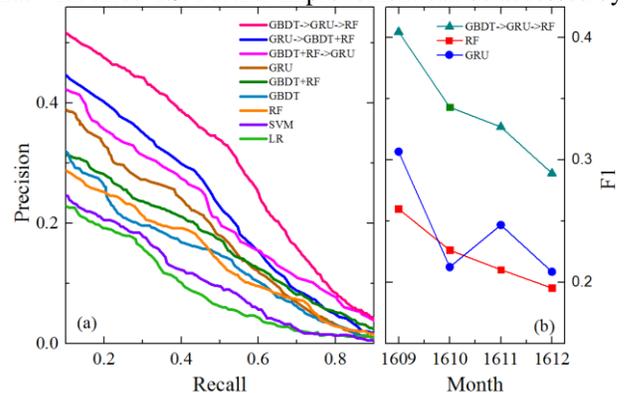

*Fig. 7. (a) Comparison of PR curves in special imbalance ratio.*
*(b) Comparison of variation trends for F1 score as time elapses*

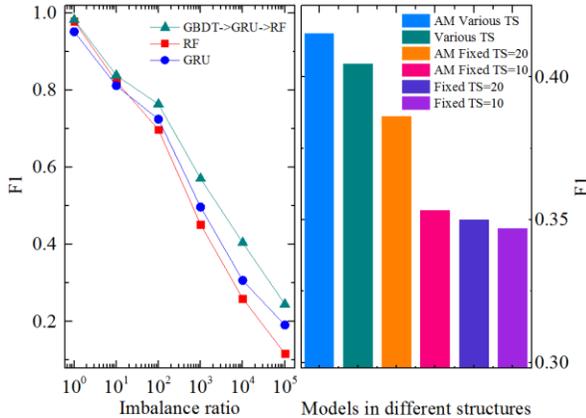

Fig. 8. (a) Comparison of decline trends for F1 score with increasing imbalance ratio. (b) Comparison of F1 score for WBW models with different GRU structures.

In summary, "WBW" sequence learning architecture provides a better performance than that of "WWB" or "BWW" structures for our business scenario, let alone other simpler structures like "WB" or "BW". In addition, model performance can be further enhanced by using various TS ranges and attention mechanism. Our future work will be focused on combining the intelligent model into production system which requires both concurrency and timeliness.

## V. CONCLUSIONS

In this paper, we presented a sophisticated solution to build a transaction fraud detection model. Firstly, artificial feature engineering work is carried out on Spark. Next, GBDT model is involved to optimize features within a single transaction. Then GRU model is applied on transformed sequential samples to learn relationships between transactions better. Finally, a top-layer RF classifier is trained using optimized transaction eigenvectors. This approach has been proved to be more efficient for detecting transaction fraud than most traditional methods. In addition, attention mechanism has also been involved for enhancing model performance. The entire collaboration model by stacking an ensemble model, a RNN deep learning model and then another ensemble model orderly can be unified as WBW sandwich-structured sequence learning architecture. Models in similar structures could also play important roles in many other scenarios, where the information sequence is made up of vectors with complex interconnected features.

ACKNOWLEDGMENT

This research was supported by National Engineering Laboratory for Electronic Commerce & Electronic Payment and FinTech Research Center in People's Bank of China, sponsored by High-Tech Service Industry R&D and Industrialization Project of National Development & Reform Commission ([2015] 289), Shanghai Sailing Program 17YF1425800 and Pudong New District Science & Technology Development Postdoctoral Fund.

REFERENCES

[1] HSN Consultants, Inc., "The Nilson Report" [Online PDF], available from https://nilsonreport.com/upload/content_promo/The_Nilson_Report_Issue_1118.pdf

[2] C. Phua, V. Lee, K . Smith and R. Gryler, "A Comprehensive Survey of Data Mining-based Fraud Detection Research," Artificial Intelligence Review, abs/1009.6119, 2010.

[3] M. I. Alowais and L. K. Soon, "Credit Card Fraud Detection: Personalized or Aggregated Model," Ftra International Conference on Mobile, Ubiquitous, and Intelligent Computing, 2012:114-119.

[4] F. T. Liu, K. M. Ting, Z. H. Zhou, "Isolation-Based Anomaly Detection," Acm Transactions on Knowledge Discovery from Data, 6(1):1-39, 2012.

[5] A. Srivastava, A. Kundu, S. Sural and A. K. Majumdar, "Credit Card Fraud Detection Using Hidden Markov Model," IEEE Transactions on Dependable & Secure Computing, 5(1):37-48, 2008.

[6] R. J. Bolton and J. H. David, "Unsupervised Profiling Methods for Fraud Detection," Proc Credit Scoring & Credit Control 7: 5-7, 2001.

[7] Z. C. Lipton, J. Berkowitz and C. A. Elkan, "Critical Review of Recurrent Neural Networks for Sequence Learning," Computer Science, 2015.

[8] B . Wiese and C. Omlin, "Credit Card Transactions, Fraud Detection, and Machine Learning: Modelling Time with LSTM Recurrent Neural Networks," 2009.

[9] V. V. Vlasselaer, C. Bravo, O. Caelen, et al., "APATE: A novel approach for automated credit card transaction fraud detection using network-based extensions," Decision Support Systems, 75:38-48, 2015.

[10] X. He, J. Pan, O. Jin, et al., "Practical Lessons from Predicting Clicks on Ads at Facebook," Proceedings of 20th ACM SIGKDD Conference on KnowledgeDiscovery and Data Mining. 2014:1-9.

[11] W. Byeon, T. M. Breuel, F. Raue, et al., "Scene labeling with LSTM recurrent neural networks," Computer Vision and Pattern Recognition IEEE, 2015:3547-3555.

[12] K. Cho, B. V. Merrienboer, D. Bahdanau, et al., "On the Properties of Neural Machine Translation: Encoder-Decoder Approaches," Computer Science, 2014.

[13] H. T. Cheng, L. Koc, J. Harmsen, et al., "Wide & Deep Learning for Recommender Systems," The Workshop on Deep Learning for Recommender Systems. ACM, 2016:7-10.

[14] B . Athiwaratkun, J. W. Stokes, "Malware classification with LSTM and GRU language models and a character-level CNN," IEEE International Conference on Acoustics, Speech and Signal Processing, 2017:2482-2486.

[15] K. Irie, Z. Tüske, T. Alkhouli, et al., "LSTM, GRU, Highway and a Bit of Attention: An Empirical Overview for Language Modeling in Speech Recognition," Interspeech, 2016:3519-3523.

[16] J. Chen, H. Zhang, X. He, et al., "Attentive Collaborative Filtering: Multimedia Recommendation with Item- and Component-Level Attention," International ACM SIGIR Conference on Research and Development in Information Retrieval, 2017:335-344.